\documentclass[%
 reprint,
runinaddress,
nofootinbib,
 amsmath,amssymb,
 aps,
pra,
]{revtex4-1}
\usepackage[dvipsnames]{xcolor}
\usepackage[english]{babel}
\usepackage{hyperref}
\usepackage{amsfonts}
\usepackage{ulem}

\makeatletter
\def\bbl@set@language#1{%
  \edef\languagename{%
    \ifnum\escapechar=\expandafter`\string#1\@empty
    \else\string#1\@empty\fi}%
  \@ifundefined{babel@language@alias@\languagename}{}{%
    \edef\languagename{\@nameuse{babel@language@alias@\languagename}}%
  }%
  \select@language{\languagename}%
  \expandafter\ifx\csname date\languagename\endcsname\relax\else
    \if@filesw
      \protected@write\@auxout{}{\string\select@language{\languagename}}%
      \bbl@for\bbl@tempa\BabelContentsFiles{%
        \addtocontents{\bbl@tempa}{\xstring\select@language{\languagename}}}%
      \bbl@usehooks{write}{}%
    \fi
  \fi}
\newcommand{\DeclareLanguageAlias}[2]{%
  \global\@namedef{babel@language@alias@#1}{#2}%
}
\makeatother

\DeclareLanguageAlias{en}{english}

\usepackage{soul}
\usepackage{graphicx}
\usepackage{amsmath}
\usepackage{array}
\usepackage{amssymb}
\usepackage{dcolumn}
\usepackage{bm}

\usepackage{mathabx}
\usepackage{enumitem} 
\usepackage{tabularx}
\usepackage{soul}

\newcommand\blfootnote[1]{%
  \begingroup
  \renewcommand\thefootnote{}\footnote{#1}%
  \addtocounter{footnote}{-1}%
  \endgroup
}

\renewcommand{\vec}[1]{\mathbf{#1}}

\def\vec#1{\boldsymbol{#1}}

\def\pd2v#1#2#3{\frac{\partial^2 #1}{\partial #2 \partial #3}}

\def \vec#1{\mathbf{#1}}
\def \2x2mat#1#2#3#4{
\left( \begin{array}{cc}
#1 &  #2 \\  #3 &  #4
\end{array} \right)
}

\begin{document}

\preprint{APS/123-QED}

\title{Pixel super-resolution with spatially-entangled photons}

\author{Hugo Defienne$^{\,1,^*}$}

\author{Patrick Cameron$^{\,1}$}%

\author{Bienvenu Ndagano$^{\,1}$}%

\author{Ashley Lyons$^{\,1}$}%

\author{Matthew Reichert$^{\,2}$}%

\author{Jiuxuan Zhao$^{\,3}$}%

\author{{Andrew R. Harvey$^{\,1}$}}%

\author{Edoardo Charbon$^{\,3}$}%

\author{Jason W. Fleischer$^{\,2}$}%

\author{Daniele Faccio$^{\,1,^\dagger}$}%

\affiliation{ \\ $^{1}$School of Physics and Astronomy, University of Glasgow, Glasgow G12 8QQ, UK \\ $^{2}$Department of Electrical and Computer Engineering, Princeton University, Princeton, USA\\ $^{3}$Advanced Quantum Architecture Laboratory (AQUA), Ecole Polytechnique Federale de Lausanne (EPFL), 2002 Neuchatel, Switzerland  \
}%

\begin{abstract}
Pixelation occurs in many imaging systems and limits the spatial resolution of the acquired images. This effect is notably present in quantum imaging experiments with correlated photons in which the number of pixels used to detect coincidences is often limited by the sensor technology or the acquisition speed. Here, we introduce a pixel super-resolution technique based on measuring the full spatially-resolved joint probability distribution (JPD) of spatially-entangled photons. Without shifting optical elements or using prior information, our technique increases the pixel resolution of the imaging system by a factor two and enables retrieval of spatial information lost due to undersampling. We demonstrate its use in various quantum imaging protocols using photon pairs, including quantum illumination, entanglement-enabled quantum holography, and in a full-field version of N00N-state quantum holography. The JPD pixel super-resolution technique can benefit any full-field imaging system limited by the sensor spatial resolution, including all already established and future photon-correlation-based quantum imaging schemes, bringing these techniques closer to real-world applications. 
\end{abstract}
\maketitle
\blfootnote{Corresponding author: \\
$^*$ hugo.defienne@glasgow.ac.uk \\
$^\dagger$ daniele.faccio@glasgow.ac.uk  }
The acquisition of a high-resolution image over a large field of view is essential in most optical imaging applications. In this respect, the widespread development of digital cameras made of millions of pixels has strongly contributed to create imaging systems with large space-bandwidth products. In classical imaging, it is therefore mainly the imaging systems with strong spatial constraints that suffer from pixelation and undersampling, such as lensless on-chip microscopes~\cite{bishara_lensfree_2010,bishara_holographic_2011}. However, this effect is very present in quantum imaging schemes where it severely hinders the progress of these techniques towards practical applications.\\
Quantum imaging systems harness quantum properties of light and their interaction with the environment to go beyond the limit of classical imaging or to implement unique imaging modalities~\cite{moreau_imaging_2019-1}. Of the many approaches, imaging schemes based on entangled photon pairs  {are the most common and are} among the most promising. Proof-of-principle demonstrations range from improving optical resolution~\cite{boto_quantum_2000,reichert_massively_2018} and imaging sensitivity~\cite{ono_entanglement-enhanced_2013,israel_supersensitive_2014,brida_experimental_2010,icfo_2021} to the creation of new imaging protocols, such as ghost imaging~\cite{pittman_optical_1995,aspden_epr-based_2013}, quantum illumination~\cite{lopaeva_experimental_2013,defienne_quantum_2019,gregory_imaging_2020,padgett_2021} and quantum holography~\cite{devaux_quantum_2019,defienne_polarization_2021-1}. Contrary to classical imaging, photon-correlation-based imaging systems operate by measuring photon coincidences between many spatial positions of the image plane in parallel (except from induced-coherence imaging approaches~\cite{lemos_quantum_2014,kviatkovsky_microscopy_2020,vanselow_frequency-domain_2020}). In practice, this process is much more delicate than forming an intensity image by photon accumulation and therefore requires specific photodetection devices. Although such a task was originally performed with raster-scanning single-photon detectors~\cite{pittman_optical_1995}, today most implementations use single-photon sensitive cameras such as electron multiplied charge coupled devices (EMCCD)~\cite{moreau_realization_2012,edgar_imaging_2012}, intensified complementary metal-oxide-semiconductor (iCMOS)~\cite{chrapkiewicz_high-fidelity_2014} and single-photon avalanche diode (SPAD) cameras~\cite{unternahrer_coincidence_2016}. \\
\begin{figure*}
\includegraphics[width=1 \textwidth]{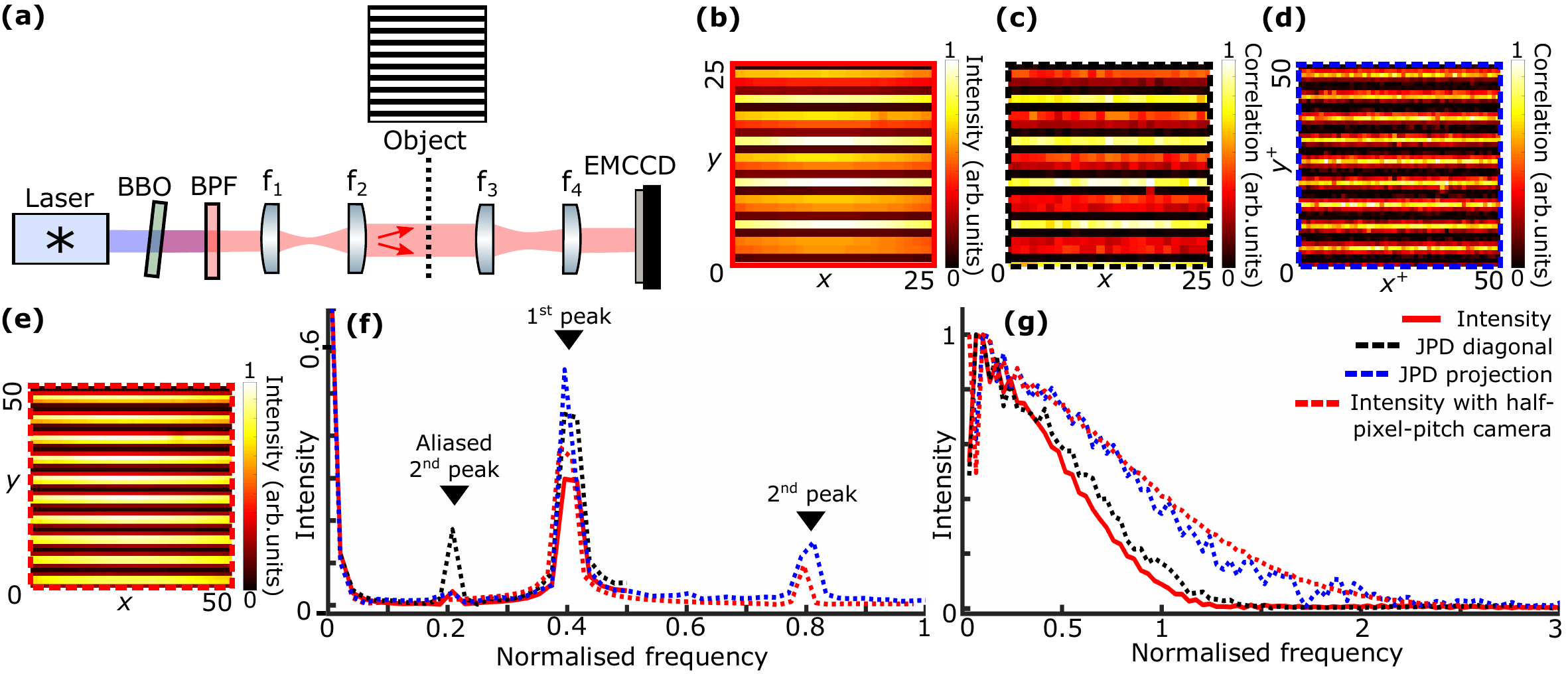} \caption{\textbf{Experimental demonstration of JPD pixel super-resolution.} \textbf{(a)} Experimental setup. Light emitted by a diode laser (405 nm) illuminates $\beta$-barium borate (BBO) crystal with a thickness of $0.5$ mm to produce spatially-entangled pairs of photons by type I SPDC. Long-pass and band-pass filters at $810 \pm 5$nm (BPF) positioned after the crystal remove pump photons. A two-lens system $f_1-f_2$ images the crystal surface onto an object $t$, that is itself imaged onto the EMCCD camera by another two-lens system $f_3-f_4$. $t$ is a grid-shaped amplitude object. {Photon correlation width in the image plane is estimated as $\sigma \approx 13 \mu$m and camera pixel pitch is $\Delta=32 \mu$m. \textbf{(b)} Intensity image, \textbf{(c)} JPD diagonal image and \textbf{(d)} sum-coordinate projection of the JPD. \textbf{(e)} Intensity image obtained using a camera with half pixel pitch i.e. $16 \mu$m. All images show the same spatial region of the object containing $10$ grating periods. Coordinates are in pixels. \textbf{(f)} Spectra of the intensity image (solid red), diagonal image (dashed black), JPD sum-coordinate image (dashed blue) and intensity image acquired with a $16 \mu$m-pixel-pitch camera (dashed red) obtained by performing a discrete Fourier transform to the corresponding image and averaging over the $x$-axis. \textbf{(g)} System modulation transfer function (MTF) obtained using the slanted-edge technique with an intensity image (solid red), a diagonal image (dashed black), a JPD sum-coordinate image (dashed blue) and an intensity image acquired using a $16 \mu$m-pixel-pitch camera (dashed red). All frequency values are normalized to the same reference frequency $k_0 = 1/\Delta$.}
\label{Figure1}}
\end{figure*}
EMCCD cameras are the most widely used devices for imaging photon correlations thanks to their high quantum efficiency, low noise and high pixel resolution. However, these cameras suffer from a very low frame rate ($\sim$100 fps) due to their electronic amplification process operating in series and therefore require very long acquisition times ($\sim$10 hours) to reconstruct correlation images~\cite{toninelli_resolution-enhanced_2019,defienne_quantum_2019,gregory_imaging_2020,defienne_polarization_2021-1}. Using a smaller sensor area or a binning technique can reduce the acquisition time, but at the cost of a loss in pixel resolution. By using an image intensifier, iCMOS cameras do not use such slow electronic amplification processes and can therefore reach higher frame rates ($\sim$1000 fps). However, so far these cameras have only enabled correlation images with relatively small number of pixels (and still several hours of acquisition), mostly because of their low detection efficiency and higher noise level~\cite{chrapkiewicz_high-fidelity_2014,chrapkiewicz_hologram_2016-1}. Finally, SPAD cameras are an emerging technology that can detect single photons at low noise while operating at very high frame rate (up to $800$ kfps) despite their low quantum efficiency ($\sim$20$\%$) and typically low resolution ($\sim$1000 pixels). \\
It is clear from the above that nearly all sensor technologies currently used in  quantum imaging experiments suffer from a poor pixel resolution, either directly when the technology does not have cameras with enough pixels or indirectly when the experiment can only operate in a reasonable time with a small number of pixels. In these systems, objects are therefore often undersampled, resulting in a loss of spatial information and the creation of artefacts in the retrieved images. 
\begin{figure*}
\includegraphics[width=1 \textwidth]{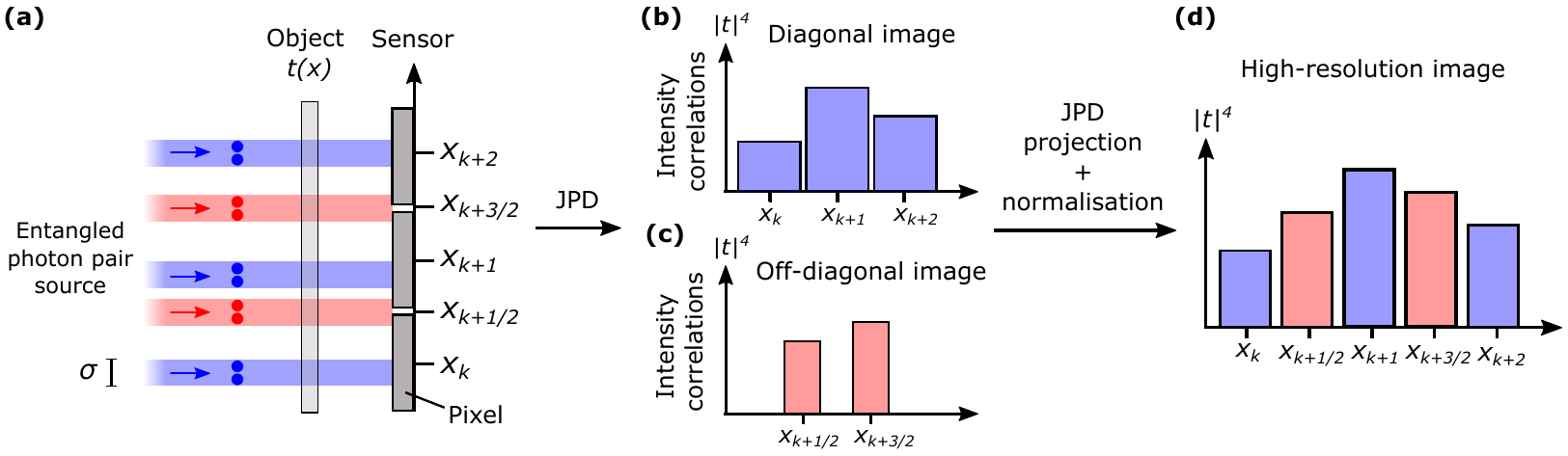} \caption{\textbf{Principle of JPD pixel super-resolution.} \textbf{(a)} Schematic of the optical imaging system composed of an object $t$ illuminated by entangled photon pairs with correlation width $\sigma$ and imaged onto an array of pixels. $\Delta$ and $\delta$ are pixel pitch and gap, respectively. Positions at the center of pixels are noted using an integer indices i.e. $x_k$, while those between pixels are noted using half-integer indices i.e. $x_{k+1/2}$. Blue rays represent some photon pairs falling on a single pixel and red rays some falling on two nearest-neighbor pixels. \textbf{(b)} and \textbf{(c)}, Unidimensional images formed by JPD diagonal elements $\Gamma_{kk}$ and off-diagonal elements $\Gamma_{kk+1}$, respectively. \textbf{(d)} Performing a sum-coordinate projection of the JPD using equation~\ref{equ2} recombines these two low-resolution images into a high-resolution one.
\label{Figure1bis}}
\end{figure*}
Here, we demonstrate a quantum image processing technique based on entangled photon pairs that increases the pixel resolution by a factor two.
We experimentally demonstrate it in three common photon-pair-based imaging schemes: two quantum illumination protocols using (i) a near-field illumination configuration with an EMCCD camera~\cite{defienne_quantum_2019} and a (ii) far-field configuration with a SPAD camera~\cite{defienne_full-field_2021-1}, and (iii) an entanglement-enabled holography system~\cite{defienne_polarization_2021-1}. In addition, we use our JPD pixel super-resolution method in (iv) a full-field version of N00N-state quantum holography, a scheme that has only been demonstrated so far using a scanning approach~\cite{ono_entanglement-enhanced_2013}. Note that we refer to our technique as 'pixel super-resolution' to avoid confusion with the term 'super-resolution' describing imaging techniques capable of overcoming the classical diffraction limit.

\section*{Results}
\noindent \textbf{{Experimental demonstration}}. Figure~\ref{Figure1}.a describes the experimental setup used to demonstrate the principle of our technique for the widely used case of {$p=2$} spatially entangled photons. The spatially entangled photon pairs are produced by type-I SPDC in a thin $\beta$-barium Borate (BBO) crystal and illuminate an object, $t$, using a near-field illumination configuration (i.e. the output surface of the crystal is imaged onto the object). {The biphoton correlation width in the camera plane is estimated to be $\sigma \approx 13\mu$m}. The object is imaged onto an EMCCD camera {with a pixel pitch of $\Delta=32\mu$m}. In our experiment, $t$ is a horizontal square-modulation amplitude grating. The recorded intensity image in Figure~\ref{Figure1}.b {shows a region of the object composed of $10$ grating periods imaged with $25$ rows of pixels. It is clear that the image suffers from the effect of aliasing, due to an harmonic above the Nyquist frequency of the detector array, leading to a low-frequency Moire modulation with a period of approximately $5$ pixels. If we are able to double the sampling frequency so as to super-resolve the image, it is expected that we will be able to image all the harmonics and hence remove this Moire pattern}. \\
We also measure the spatially-resolved JPD of the photon pairs {$\Gamma_{ijkl}$ by identifying photon coincidences between any arbitrary pair of pixels $(i,j)$ and $(k,l)$ centred at spatial positions $\vec{r_1}=(x_i,y_j)$ and $\vec{r_2}=(x_k,y_l)$} using the method described in~\cite{defienne_general_2018-2}. The information contained in the JPD can be used for various purposes. For example, it was used in pioneering works with EMCCD cameras to estimate position and momentum correlation widths of entangled photons pairs in direct imaging of the Einstein-Podolsky-Rosen paradox~\cite{moreau_realization_2012,edgar_imaging_2012}. In the experiment here, the diagonal component of the JPD, {$\Gamma_{ijij}$} reconstructs an image of the object {i.e. $\Gamma_{ijij} = |t(x_i,y_j)| ^4$}. Such a diagonal image is the quantity that is conventionally measured and used in all photon-pair-based imaging schemes using a near-field illumination configuration~\cite{reichert_biphoton_2017,reichert_massively_2018,toninelli_resolution-enhanced_2019,defienne_quantum_2019}. As shown in Figure~\ref{Figure1}.c, measuring the diagonal image in our experiment does not however improve the image quality i.e. the Moire pattern is still present.  \\
\begin{figure*}
\includegraphics[width=0.95 \textwidth]{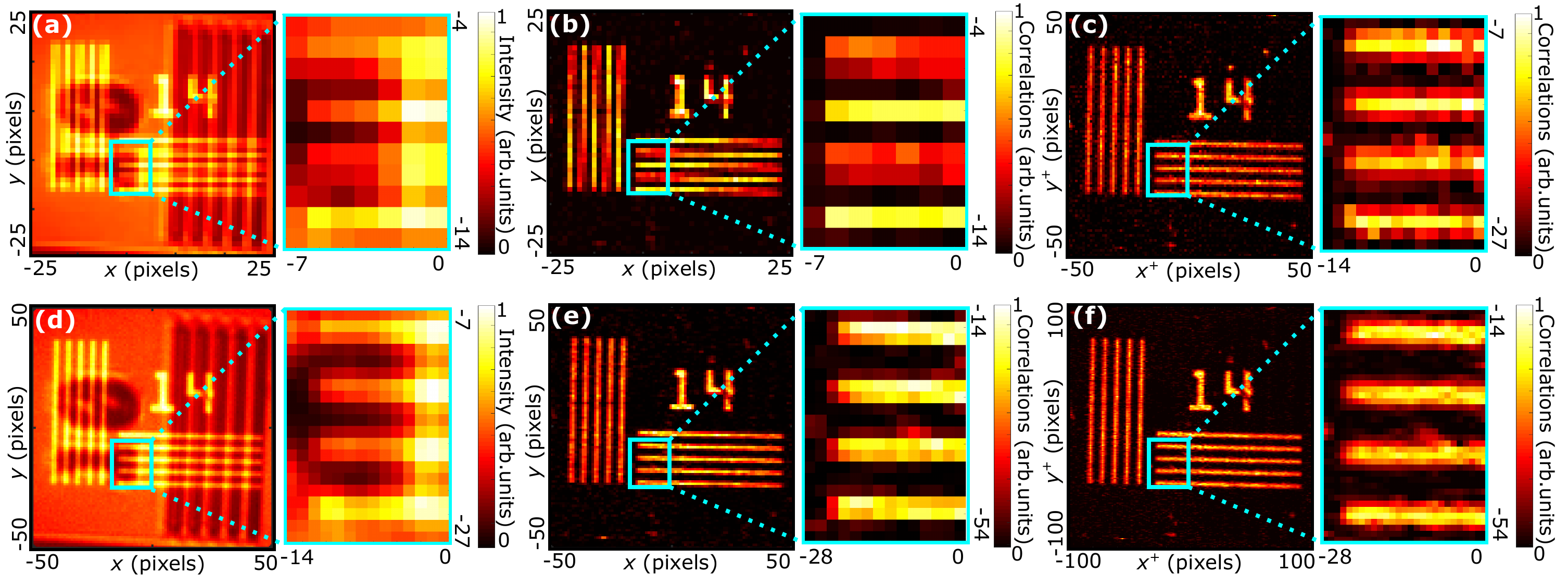} \caption{{\textbf{Results in quantum illumination imaging}. Experiment performed using a setup similar to this shown in Figure~\ref{Figure1}.a. in which another object illuminated by a classical light source (LED) is inserted. Images of both objects (positive and negative resolution charts) are super-imposed onto the camera. \textbf{(a)} Intensity image, {\textbf{(b)} diagonal image and} \textbf{(c)} {{JPD pixel super-resolution image}} acquired using a camera with a $32$ $\mu$m-pixel-pitch. \textbf{(d)} Intensity image, {\textbf{(e)} diagonal image and} \textbf{(f)} JPD pixel super-resolution image acquired using a camera with a $16$ $\mu$m-pixel-pitch. The photon correlation width in the camera plane is $\sigma\approx 8$ $\mu$m. }\\
\label{Figure2}}
\end{figure*}
There is another way to retrieve an image of the object without using the diagonal component of the JPD. For that, one can project the JPD along its sum-coordinate axis, achieved by {summing} $\Gamma$: {
	\begin{equation}
	\label{equ2}
	P^+_{i^+j^+} = \sum_{i=1}^{N_X} \sum_{j=1}^{N_Y} \Gamma_{(i^+-i) \,(j^+-j) \, i\, j} \, ,
	\end{equation}}
where {$N_Y \times N_X$ is the number of pixels of the illuminated region of the camera sensor}, $P^+$ is defined as the sum-coordinate projection of the JPD and $(i^+,j^+)$ are sum-coordinate pixel indexes. Such a projection retrieves an image of the object sampled over four times the number of pixels of the sensor: $P^+_{i^+j^+} = |t(x_i,y_j)| ^4$ for even pixel indices i.e. $i^+ = 2i$ and $j^+ = 2j$, with $x_i = i \Delta$ and $x_j = j \Delta$ ; $P^+_{i^+j^+} = |t(x_{i+1/2},y_{j+1/2})| ^4$ for odd pixel indices i.e. $i^+ = 2i+1$ and $j^+ = 2j+1$, with $x_{i+1/2} = x_i+\Delta/2$ and $y_j+\Delta/2$. Pixel resolution is therefore increased by a factor $2$ (see section \textit{JPD pixel super-resolution principle} for more details). Figure~\ref{Figure1}.d shows the resulting sum-coordinate projection of the JPD measured in the experiment shown in Figure~\ref{Figure1}.a. Thanks to pixel super-resolution, we observe that the spurious low frequency Moire modulation has been removed and the 10 grating periods are now clearly visible. As a comparison, such a high-resolution image is very similar to a conventional intensity image acquired using a camera with half the pixel pitch i.e. $16\mu$m (Fig.~\ref{Figure1}.e). \\
The frequency analysis of these different images is shown in Figure~\ref{Figure1}.f and provides more quantitative information about our approach. In particular, we observe the absence of a frequency peak at $0.2$ in the sum-coordinate image spectrum (dashed blue), while it is present in both intensity (solid red) and JPD-diagonal (dashed black) image spectra. It is instead substituted by a peak at $0.8$ that is the true frequency component of the object (harmonics), as confirmed by its presence in the spectrum of the intensity image acquired using the high-resolution, $16\mu$m-pixel-pitch camera (dashed red). Removal of the aliased frequency component corresponds also to the disappearance of the Moire pattern in the real space. This confirms that our approach achieves pixel super-resolution and retrieves information that was lost due to under-sampling. Additional measurements provided in supplementary document section 3 are acquired using a camera of even lower resolution ($48\mu$m-pixel-pitch) and show that JPD pixel super-resolution can also recovers the fundamental spectral component (main peak) even when this is absent in both the intensity and diagonal images. JPD pixel super-resolution is also confirmed by simulations detailed in supplementary document section 2.8.  \\
{Figure~\ref{Figure1}.g shows system modulation transfer functions (MTF) calculated using the slanted-edge technique~\cite{burns2000slanted} with different imaging modalities. The MTF obtained using the JPD sum-coordinate projection (dashed blue) is approximately $1.7$ times broader than those acquired using intensity (solid red) and JPD diagonal (dashed black) images, almost matching the MTF retrieved by intensity measurement with the high-resolution $16\mu$m-pixel-pitch camera (dashed red). This shows that JPD pixel super-resolution not only doubles the Nyquist frequency ($1/(2 \Delta) \rightarrow 1/\Delta$), but also broadens the system MTF which results in less attenuation of higher frequencies. The broadening of the MTF is explained by the fact that the effective size of a 'pixel' during a JPD measurement (i.e. the size of the surface over which the coincidences are integrated) is on average smaller than the real size of the pixels. This shows that images retrieved by JPD pixel super-resolution are similar to conventional intensity images obtained with a camera that has 4 times more pixels (higher Nyquist frequency) but that also has smaller pixels (broader MTF) (more details about MTF measurements are provided in Methods and in supplementary document section 4.)} \\
Interestingly, results in Fig.~\ref{Figure1} also lead to another conclusion that, contrary to a common belief in the field, conventional imaging with photon pairs (i.e. using the JPD diagonal) does not always improve image quality compared to classical intensity imaging. For example here, the spurious Moire effect in the diagonal image (Fig.~\ref{Figure1}.c) is more intense than in the direct intensity image (Fig.~\ref{Figure1}.b), which is also confirmed by a higher intensity of the $0.2$ frequency peak in the diagonal image spectra (Fig.~\ref{Figure1}.f).\\
{\noindent \textbf{JPD pixel super-resolution principle}.} We gain further insight into the underlying principle of JPD pixel super-resolution from inspection of the JPD of a spatially entangled two-photon state: this is a $4$-dimensional object containing much richer information than a conventional 2D intensity image. The JPD contains correlation information not only between photons detected at the same pixel, but also correlation information about photons detected between nearest-neighbour pixels.\\
{Figure~\ref{Figure1bis} illustrates this concept in the one-dimensional case. In Fig.~\ref{Figure1bis}.a, photon pairs with a correlation width $\sigma$ illuminate a one-dimensional object $t(x)$ imaged onto an array of pixels with pitch $\Delta$ and pixel gap $\delta$ {{ (i.e. the width of non-active areas between neighbouring pixels)}. When measuring the JPD, there are two main contributions: (i) The first contribution originates from pairs of photons detected at the same pixel (blue rays) and form the JPD diagonal elements, $\Gamma_{kk}$. Because these pairs crossed the object around the same positions as the pixels i.e. $x_k$, the resulting image (Fig.~\ref{Figure1bis}.b) provides a sampling of the object $\Gamma_{kk} \sim |t(x_k)|^4$ similar to that performed by a conventional intensity measurement. (ii) The second contribution originates from photon pairs detected by nearest-neighbour pixels and forms the JPD off-diagonal elements $\Gamma_{kk+1}$. Because these photons crossed the object around positions located between the pixels i.e. $x_{k+1/2}$, the resulting image (Fig.~\ref{Figure1bis}.c) provides a sampling of the object $\Gamma_{kk+1} \sim |t(x_{k+1/2})|^4 $ similar to an intensity measurement performed with a sensor shifted by $\Delta/2$ in the transverse direction (see Methods for derivations of $\Gamma_{kk}$ and $\Gamma_{kk+1}$). \\
Finally, projecting the JPD along the sum-coordinate (Eq.~\ref{equ2}) retrieves a high-resolution image (Fig.~\ref{Figure1bis}.d) by interlacing diagonal and off-diagonal elements. To understand this recombination, one can expand Eq.~\ref{equ2} in the one dimensional case for even and odd pixels :
\begin{eqnarray}
P^+_{i^+=2k} &=& \Gamma_{kk} + 2 \sum_{l=0}^{N} \Gamma_{k-2l k} \label{equA} \\
P^+_{i^+=2k+1} &=& 2 \Gamma_{kk+1} + 2 \sum_{l=1}^{N} \Gamma_{k-(2l+1) k} \label{equB} 
\end{eqnarray}
where $N$ is the number of pixel of the sensor. In theory, when operating under the constraint $\sigma < \Delta$, correlations between non-neighbouring pixels are nearly zeros i.e. $\Gamma_{ij} \approx 0$ if $|i-j|>2$. The sum terms in Eqs.~\ref{equA} and~\ref{equB} then become negligible compared to $\Gamma_{kk}$ and $\Gamma_{kk+1}$. It leads to $P^+_{i^+=2k} = \Gamma_{kk}$ and $P^+_{i^+=2k+1} = 2 \Gamma_{kk+1}$, providing the super-resolved image. In practice, however, experimentally measured correlation values are noisy. All the noise then adds up in the sums in Eqs.~\ref{equA} and~\ref{equB}, ultimately producing a noise with greater variance that dominates the diagonal and off-diagonal elements in the final image. To solve this issue, the JPD is filtered before performing the projection to remove the weakest correlation values i.e. all terms except $\Gamma_{kk}$ and $\Gamma_{kk+1}$ are set to zero. In doing so, we also remove the noise associated with these values, that do not add up in the sums, and significantly improve the quality of the final image (see Methods and Figure~\ref{Figure5} for more details). \\
In addition, it should be noted that JPD diagonal values (i.e. $\Gamma_{ijij}$ in the two dimensional case) are not measured directly with single-photon cameras, as most of these devices are not photon-number resolving. To circumvent this limitation, diagonal values are estimated from correlations between either vertical ($\Gamma_{i(j \pm 1)ij}$) or horizontal ($\Gamma_{(i \pm 1)jij}$) direct-neighbouring pixels. This has the practical consequence of restricting the super-resolution effect to one dimension in the complementary axis. In the general case, an image super-resolved in two dimensions can still be obtained by combining two images super-resolved in each direction. In the specific case of EMCCD cameras, this is however not possible because of charge smearing~\cite{reichert_massively_2018}. This effect compromises horizontal correlations values and truly restricts the super-resolution effect to one dimension along the vertical spatial axis (see Methods). These limitations are lifted when operating in a far-field illumination configuration~\cite{gregory_imaging_2020,defienne_polarization_2021-1,defienne_full-field_2021-1} (see section \textit{Application to quantum holography}) and in quantum imaging schemes using two distinct cameras~\cite{devaux_quantum_2019,devaux_imaging_2020}.\\
The previous analysis also shows that two experimental conditions must be verified to achieve pixel super-resolution. First, the sensor fill-factor must be large enough to allow photon coincidence detection between neighbouring pixels i.e. $\delta < \sigma$. Second, the spatial resolution of the imaging system must be limited by the pixel size. This is true if the higher spatial frequency component of the optical field in the object plane is both smaller than the imaging system spatial frequency cut-off and larger than the sensor Nyquist frequency $1/(2\Delta)$. In practice, one must verify at least that $\sigma < \Delta$. Otherwise, a JPD sum-coordinate projection can still be retrieved, but will not contain more information than a conventional intensity measurement. In the following, we apply our approach to real quantum imaging experiments in which the condition $\delta < \sigma < \Delta$ is true. {{We note that this condition is  straightforward to satisfy. Indeed, $\sigma < \Delta$ is the starting requirement for any form of pixel super-resolution to actually make sense - without this condition, there will not even be any aliasing effects in the first place. The other condition $\delta < \sigma$, is always satisfied unless the pixel fill factor is extremely low. }}} 
\begin{figure}
\includegraphics[width=1 \columnwidth]{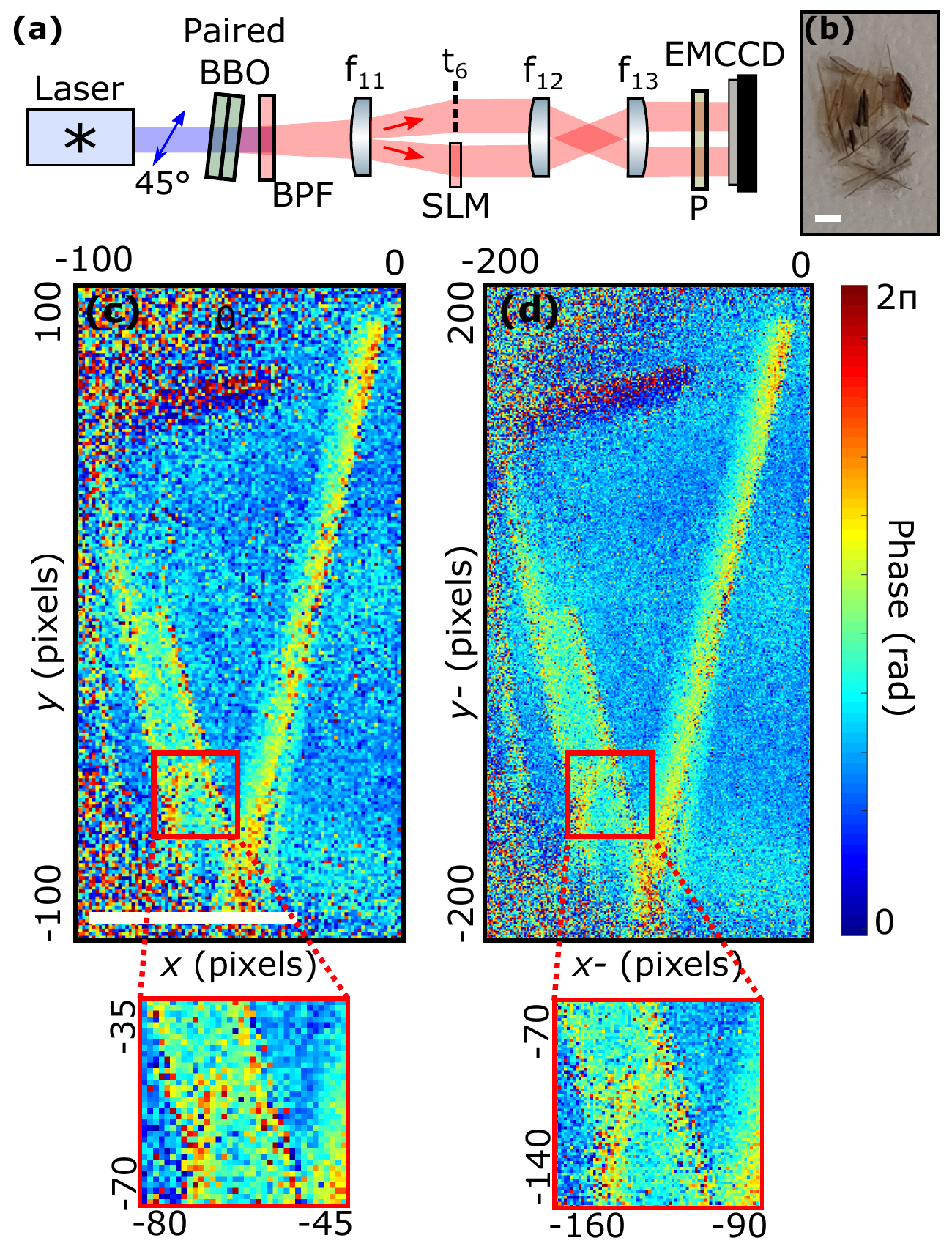} \caption{ \textbf{Results in entanglement-enabled quantum holography}.\textbf{(a)} Experimental setup. Light emitted by a laser diode at $405$ nm and polarized at 45$^{\circ}$ illuminates a stacked BBO crystal pair ($0.5$ mm thickness each) whose optical axes are perpendicular to each other to produce pairs of photons entangled in space and polarization by type-I SPDC. After the crystals, pump photons are filtered out by a band pass filter at $810 \pm 5$nm. A lens $f_{11}$ is used to Fourier-image photon pairs onto an optical plane where an SLM with a phase object $t_6$. A two-lens imaging system $f_{12}-f_{13}$ is then used to image the SLM and the object onto two different parts of an EMCCD camera. A polarizer (P) at 45$^{\circ}$ is positioned before the camera. {The biphoton correlation width in the camera plane is estimated as $\sigma \approx 9$ $\mu$m. EMCCD camera pixel pitch is $16$ $\mu$m.} \textbf{(b)} Photo of the birefringent object used in the experiment  (section of a bird feather). \textbf{(c)} Spatial phase of the object reconstructed by combining four anti-diagonal images {$\Gamma_{ij\,-i\,-j}$} measured for four phase shifts $\{0,\pi/2,\pi/2,3\pi/4 \}$ programmed by the SLM. \textbf{(d)} Spatial phase of the same object reconstructed from four minus-coordinate projections {$P^-$}. White scale bar is $1$ mm. 
\label{Figure3}}
\end{figure}
\\
\noindent \textbf{Applications to quantum illumination.} We demonstrate our technique on two different experimental schemes based on  {two-photon} quantum illumination protocols described in~\cite{defienne_quantum_2019} and~\cite{defienne_full-field_2021-1}. In the first {protocol}, an amplitude object $t_1$ is illuminated by photon pairs and imaged onto an EMCCD camera using a {similar experimental setup than this shown in Figure~\ref{Figure1}.a}. Simultaneously, another object $t_2$ is illuminated by a classical source and also imaged on the camera {(see Methods and supplementary document section 7.1 for more details about the experimental setup)}. The intensity image, Fig.~\ref{Figure2}.a, shows a superposition of both quantum and classical images. The goal {of such protocol} is to segment the quantum object and therefore retrieve an image showing only object $t_1$ illuminated by photon pairs, effectively removing any classical objects or noise. {Conventionally, this is achieved} by measuring the diagonal image shown in Figure~\ref{Figure2}.b. Using the JPD pixel super-resolution processing method, we can now perform such a task and simultaneously retrieve a pixel super-resolved image of $t_1$ (Fig.~\ref{Figure2}.c), {in which we observe the clear removal of the Moire effect due to aliasing.} \\
{In addition, we reproduced the same experiment using a higher resolution camera i.e. $16$ $\mu$m-pixel-pitch (Figs.~\ref{Figure2}.d-f). In this case, we observe that JPD pixel super-resolution anti-aliasing is mainly visible at the edges and corners of the object, in particular with the attenuation of the so-called staircase effects. Thus, even in imaging situations where aliasing does not produce artifacts as clear as Moire effect, the JPD pixel super-resolution approach still provides clear benefits in removing it and improving the overall image quality.}\\
{We also apply our technique in another quantum illumination protocol demonstrated in~\cite{defienne_full-field_2021-1}. This protocol performs the same task as in Figure~\ref{Figure2} i.e. distilling the quantum image from the classical, but uses a far-field illumination configuration, operates in reflection and detects photons with a SPAD camera (see supplementary document section 7.2 for results and experimental arrangement).}}\\
\noindent \textbf{Application to quantum holography.} We now demonstrate JPD pixel super-resolution on two different experimental quantum holography schemes. The first approach was demonstrated in~\cite{defienne_polarization_2021-1} and its experimental configuration is described in Fig.~\ref{Figure3}.a. Pairs of photons entangled in space and polarisation illuminate an SLM and a birefringent phase object $t_6$ positioned in each half of an optical plane using a far-field illumination configuration. The object and SLM are then both imaged onto two different parts of an EMCCD camera.\\
{In such a far-field configuration, the information of the JPD is now concentrated around the anti-diagonal (i.e. $\Gamma_{ij\,-i\,-j}$) because photon pairs are spatially anti-correlated in the object plane~\cite{schneeloch_introduction_2016}. This anti-diagonal is the quantity that is conventionally measured and used in all photon-pair-based imaging schemes that use a far-field illumination configuration~\cite{gregory_imaging_2020,defienne_polarization_2021-1,defienne_full-field_2021-1}. In this case, the JPD pixel super-resolution technique must be adapted by using the minus-coordinate projection $P^-$ of the filtered JPD in place of the sum-coordinate projection $P^+$ to retrieve the high-resolution image (see Methods).}\\
{The object {considered here} is a section of a bird feather, shown in Fig.~\ref{Figure3}.b}. The SLM is used to compensate for optical aberrations and implement a four phase-shifting holographic process by displaying uniform phase patterns with values $\{0,\pi/2,\pi,3\pi/2\}$ (see Methods). In the original protocol, four different images are obtained for each step of the process by measuring the anti-diagonal component $\Gamma_{ij\,-i\,-j}$ of the JPD. These images are then combined numerically to reconstruct the spatial phase of the birefringent object (Fig.~\ref{Figure3}.c). Using the new JPD pixel super-resolution approach, the four images are now obtained from the minus-coordinate projection {$P^-$} of the JPD and recombined to retrieve a phase image with improved spatial resolution (Fig.~\ref{Figure3}.d). {We do not observe the clear removal of aliasing artefacts in the super-resolved image as in Figures~\ref{Figure1}.d and~\ref{Figure2}.c due to the relatively smooth shape of the bird feather that is mostly composed of low frequencies below the Nyquist limit. Resolution improvements are therefore mainly located at the edges and corners, that visually translates into an overall improvement of the image quality.}\\
The second approach is a full-field version of a N00N-state holographic scheme based on photon pairs ($N=2$). N00N states {are known for} providing phase measurements with $N$-times better sensitivity than classical holography~\cite{israel_supersensitive_2014}. Our results show that a full-field version of N00N-state holography with photon pairs not only preserves the two-fold sensitivity enhancement, but also provides a better pixel resolution, an advantage that could only be matched by a scanning approach~\cite{ono_entanglement-enhanced_2013} that would also increase the acquisition time by a factor of four (see supplementary document section 6 for experimental layout and results.)

\section*{Discussion} We have introduced a pixel super-resolution technique based on the measurement of a spatially-resolved JPD of spatially-entangled photons. This approach retrieves spatial information about an object that is lost due to pixelation, without shifting optical elements or relying on prior knowledge. We demonstrated that this JPD pixel super-resolution approach can improve the spatial resolution in already established quantum illumination and entanglement-enabled holography schemes, as well as in a full-field version of N00N-state quantum holography, using different types of illumination (near-field and far-field) and different single-photon-sensitive cameras (EMCCD and SPAD). Our approach has the advantage that it can be used immediately in quantum imaging schemes based on photon pairs, and even in some cases by only reprocessing already acquired data. In addition, our approach can also be implemented in any classical imaging system limited by pixelation, after substituting the classical source by a source of  {correlated} photons with similar properties. Indeed, contrary to classical pixel super-resolution techniques, such as shift-and-add approaches~\cite{farsiu_fast_2004}, wavelength scanning~\cite{luo_pixel_2016} and machine-learning-based algorithms~\cite{shi_real-time_2016}, the JPD pixel super-resolution approach has the  advantage that it does not require displacing optical elements in the system or having prior knowledge about the object being imaged.  Although we experimentally demonstrated this technique for the case of $p=2$ (photon pair) entanglement, we anticipate that our approach could be generalised for all $p$ to further increase the pixel resolution. Photon pair sources are without doubt the current experimental choice in any given lab but recent efforts have shown promising progress towards direct generation of spatially entangled three-photon states~\cite{borshchevskaya_three-photon_2015}. We also underline that although the schemes shown here used spatially entangled photons, strictly speaking it is not entanglement but only spatial correlations that are used to generate the JPD. This opens the intriguing prospect for future work to investigate the potential of classical sources of light, e.g. thermal light, to achieve similar pixel super-resolution as shown here but with ready access to {$p>2$} JPDs.

\clearpage

\section*{Methods} 
\begin{figure}
\includegraphics[width=1 \columnwidth]{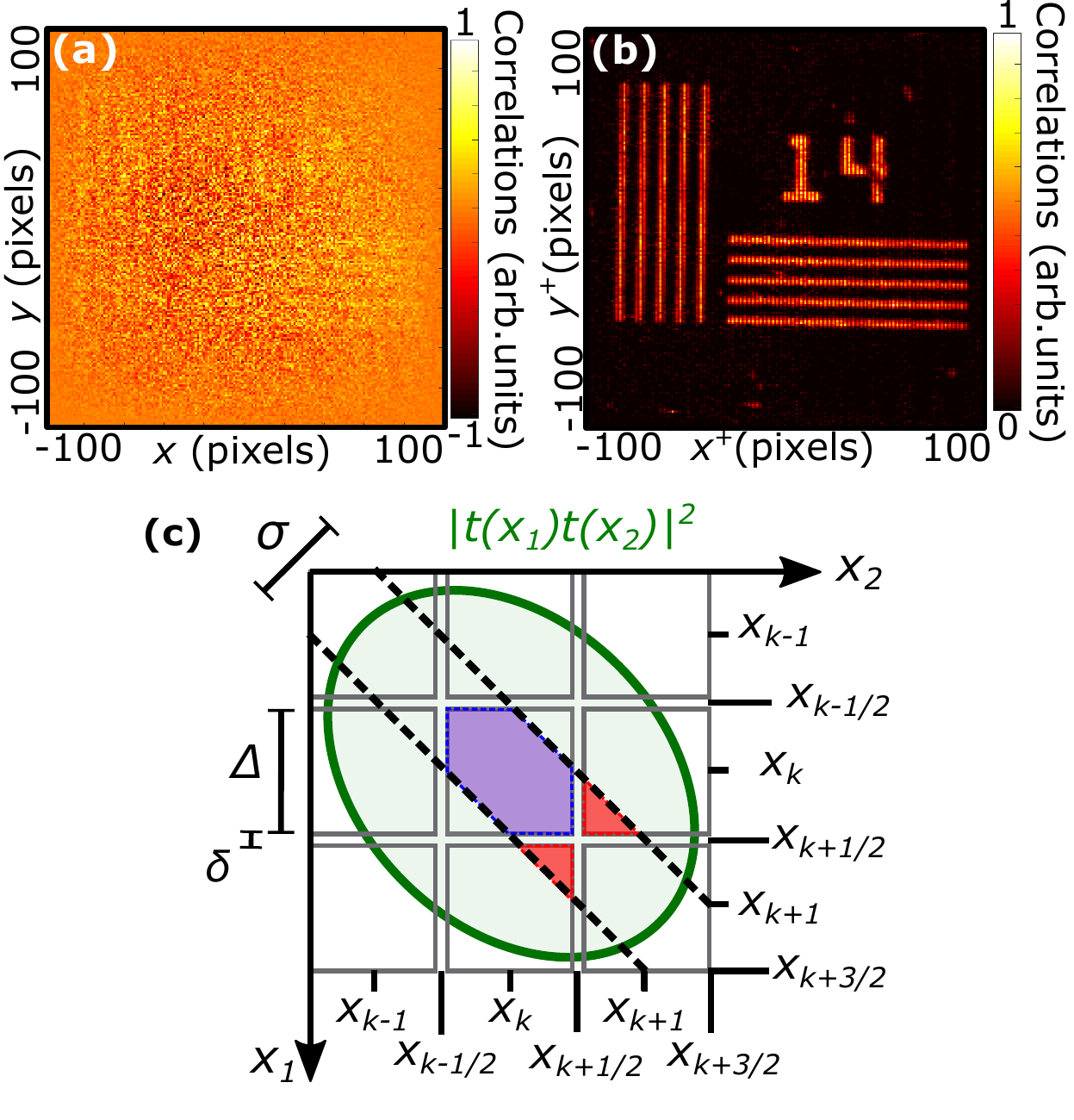} \caption{\textbf{Filtering and normalization.} \textbf{(a)} Sum-coordinate projection of the JPD $P^+$ measured in the experiment in Figure~\ref{Figure2} without filtering. \textbf{(b)} Sum-coordinate projection of the JPD $P^+$ measured in the experiment in Figure~\ref{Figure2} after filtering and before normalisation. \textbf{(c)} Graphical representation of Equation~\ref{anadev}.
\label{Figure5}}
\end{figure}

\noindent \textbf{Experimental layouts.} \textit{Experiment in Fig.~\ref{Figure1}.a:} BBO crystal has $0.5 \times 5 \times 5$ mm and is cut for type I SPDC at $405$ nm with a half opening angle of $3$ degrees (Newlight Photonics). It is slightly rotated around horizontal axis to ensure near-collinear phase matching of photons at the output (i.e. ring collapsed into a disk). The pump is a continuous-wave laser at $405$ nm (Coherent OBIS-LX) with an output power of approximately $200$ mW and a beam diameter of $0.8\pm 0.1$ mm. A $650$ nm-cut-off long-pass filter is used to block pump photons after the crystals, together with a band-pass filter centred at $810 \pm 5$ nm. The camera is an EMCCD (Andor Ixon Ultra 897) that operates at $-60^{\circ}$C, with a horizontal pixel shift readout rate of $17$ MHz, a vertical pixel shift every $0.3$ $\mu$s, a vertical clock amplitude voltage of $+4$V above the factory setting and an amplification gain set to $1000$. {The camera sensor has a total $512 \times 512$ pixels with $16\mu$m pixel pitch and nearly unity fill factor. In Figures~\ref{Figure1}.b-d, the camera is operated with a pixel pitch of $\Delta = 32$ $\mu$m by using a $2\times2$ binning. In Figure~\ref{Figure1}.e, the camera is operated with a pixel pitch of $16$ $\mu$m.} Exposure time is set to $2$ ms. The camera speed is about $100$ frames per second (fps) using a region of interest of $100 \times 100$ pixels. The two-lens imaging system $f_1-f_2$ in Fig.~\ref{Figure1}.a is represented by two lenses for clarity, but is composed of a series of $6$ lenses with focal lengths $45$ mm - $75$ mm - $50$ mm - $150$ mm - $100$ mm - $150$ mm. The first and the last lens are positioned at focal distances from the crystal and the object, respectively, and the distance between two lenses in a row equals the sum of their focal lengths. Similarly, the second two-lens imaging system $f_3-f_4$ in Figure~\ref{Figure1}.a is composed of a series of $4$ consecutive lenses with focal lengths $150$mm - $50$ mm - $75$ mm - $100$ mm arranged as in the previous case. {The imaging system magnification is $3.3$. The photon correlation width in the camera plane is estimated as $\sigma \approx 13$ $\mu$m.} \\
{\textit{Experiment used to acquire images in Figure~\ref{Figure2}:} The experimental setup is the same as this shown in Figure~\ref{Figure1}.a, with some changes in the lenses used and the addition of an external arm to superimposed the classical image. It is shown in Figure~15.a of the supplementary document. The output surface of the crystal is imaged onto an object $t_1$ using a two-lenses imaging system with focal lengths $f_5=35$ mm and $f_6=75$ mm. The object is then imaged onto the camera using a single-lens imaging system composed of one lens with focal length $f_7=50$ mm positioned at a distance of $100$mm from the object and the camera. Another object $t_2$ is inserted and illuminated by a spatially filtered light-emitting diode (LED) and spectrally filtered at $810 \pm 5$ nm. Images of both objects are superimposed on the camera using a beam splitter. $t_1$ and $t_2$ are negative and positive amplitude USAF resolution charts, respectively. They are shown in Figures~N.b and c. Exposure time is set to $6$ ms. {The imaging system magnification $2.1$. The biphoton correlation width in the camera plane is estimated as $\sigma \approx 8$ $\mu$m. $6.10^6$ frames were acquired to retrieve intensity image and JPD in approximately $20$ hours of acquisition.} The same setup was used in~\cite{defienne_quantum_2019}. {More details in section 7 of the supplementary document}\\}
\textit{Experiment in Figure~\ref{Figure3}.a:} The paired set of stacked BBO crystals have dimensions of $0.5 \times 5 \times 5$ mm each and are cut for type I SPDC at $405$ nm. They are optically contacted with one crystal rotated by $90$ degrees about the axis normal to the incidence face. Both crystals are slightly rotated around horizontal and vertical axis to ensure near-collinear phase matching of photons at the output (i.e. rings collapsed into disks). The pump laser and camera are the same than in Figure~\ref{Figure1}.a. A $650$ nm-cut-off long-pass filter is used to block pump photons after the crystals, together with a band-pass filter centred at $810 \pm 5$ nm. The SLM is a phase only modulator (Holoeye Pluto-2-NIR-015) with $1920 \times 1080$ pixels and a $8$ $\mu$m pixel pitch. For clarity, it is represented in transmission in Figure~\ref{Figure3}.a, but was operated in reflection. Exposure time is set to $3$ ms. The camera speed is about $40$ frame per second using a region of interest of $200 \times 200$ pixels. The single-lens Fourier imaging system $f_{11}$ is composed of a series of three lenses of focal lengths $45$ mm - $125$ mm - $150$ mm. The first and last lenses are positioned at focal distance from the crystal and the object-SLM plane, respectively, and the distance between each pair of lenses equals the sum of their focal lengths. The two-lens imaging system $f_{12}-f_{13}$ is in reality composed by a series of $4$ lenses with focal lengths $150$ mm - $75$ mm - $75$ mm - $100$ mm. The first and the last lens are positioned at focal distances from respectively the SLM-object and the camera, respectively, and the distance between two lenses in a row equals the sum of their focal lengths. {The imaging system effective focal length is $36$ mm.} {The photon correlation width in the camera plane is estimated as $\sigma\approx 9$ $\mu$m.} $2.5.10^6$ frames were acquired to retrieve intensity image and JPD in each case in approximately $17$ hours of acquisition. The same setup was used in~\cite{defienne_polarization_2021-1}. {More details in section 5 of the supplementary document}. \\
\\
\noindent \textbf{JPD measurement with a camera.} {$\Gamma_{ijkl}$}, {where $(i,j)$ and $(k,l)$ are two arbitrary pair of pixels centred at positions $\vec{r_1}=(x_i,y_j)$ and $\vec{r_2}=(x_k,y_l)$}, is measured by acquiring a set of $M+1$ frames $\{I^{(l)} \} _{l \in  [\![ 1,M+1]\!]}$ using a fixed exposure time and then processing them using the formula: {
\begin{equation}
\label{equ0}
\Gamma_{ijkl} = \frac{1}{M} \sum_{l=1}^M \left[  I^{(l)}_{ij}I^{(l)}_{kl} - I^{(l)}_{ij} I^{(l+1)}_{kl} \right]
\end{equation}}
In all the results of our work, $M$ was on the order of $10^6-10^7$ frames. However, it is essential to note that not all the JPD values can be directly measured using this process. When using an EMCCD camera, (i) correlation values at the same pixel, i.e. {$\Gamma_{ijij}$}, cannot be directly measured because Equation~\ref{equ0} is only valid for different pixels {$(i,j) \neq (k,l)$}~\cite{defienne_general_2018-2}, and (ii) and correlation values between vertical pixels, i.e. {$\Gamma_{iji\,(j \pm q)}$} (where $q$ is an integer that defines the position of a pixel above or bellow {$(i,j)$}), cannot be measured because of the presence of charge smearing effects. To circumvent this issue, these values are interpolated from neighbouring correlation values of the JPD i.e. {$[\Gamma_{ij\,(i+1)\,j}+\Gamma_{ij\,(i-1)\,j}] /2 \rightarrow \Gamma_{ijij}$ and $[\Gamma_{ij\,(i+1)\,(j \pm q)}+\Gamma_{ij\,(i-1)\,(j \pm q)} ] /2 \rightarrow \Gamma_{iji\,(j \pm q)}$}, as detailed in~\cite{reichert_massively_2018}. As a result, the gain in resolution along the $x$-axis in the experiments using near-field imaging configurations (Figures~\ref{Figure1} and~\ref{Figure2}, and Figure~15 and~10 of the supplementary document) is not optimal. However, it is important to note that the gain in pixel resolution along the $y$-axis is not affected by this interpolation technique. Therefore, the spectral analyses performed in Figures~\ref{Figure1} are also not impacted because the grid-objects used are horizontal (no spectral component on the $x$-axis) and all the resulting spectral curves are obtained by summing along the $x$-axis. In addition, this interpolation also does not affect experiments using far-field illumination configurations in Figure~\ref{Figure3}.a and Figure~16 of the supplementary document because the JPD diagonals do not contain any relevant imaging information (that is in the JPD anti-diagonals). More details are provided in~\cite{defienne_general_2018-2} and in section 1 of the supplementary document.\\
\\
\noindent \textbf{JPD pixel super-resolution in far-field illumination configuration.} When an object is illuminated by photon-pairs using a far-field illumination configuration (i.e. the crystal is Fourier-imaged on the object), the JPD pixel super-resolution technique must be adapted and the sum-coordinate projection $P^+$ cannot be used to retrieve the high-resolution image. {First, instead of the diagonal}, information about the object {is} retrieved by displaying the anti-diagonal component of the JPD i.e. {$\Gamma_{ij\,-i\,-j} \approx |t(x_i,y_i)|^2 |t(-x_i,-y_i)|^2$. $\Gamma_{ij\,-i\,-j}$} is the quantity that is conventionally measured and used in all photon-pairs-based imaging schemes using a far-field illumination configuration~\cite{gregory_imaging_2020,defienne_full-field_2021-1,defienne_polarization_2021-1}. {Second, instead of using the sum-coordinate projection, a super-resolved image of the object is retrieved by integrating the JPD along its minus-coordinate axis:
\begin{equation}
\label{equ3s}
P^-_{i^-j^-} = \sum_{i=1}^{N_X} \sum_{j=1}^{N_Y} \Gamma_{(i^- +i) \,(j^- +j) \, i\, j} \, ,
\end{equation}
}where {$N_Y \times N_X$ is the number of pixels of the illuminated region of the camera sensor}, $P^-$ is defined as the minus-coordinate projection of the JPD and {$(i^-,j^-)$ {defines} the sum-coordinate pixel to which a spatial variable $\vec{r^-}=(x_{i^-},y_{j^-})$} is associated. Each value {$P^-_{i^-j^-}$} is obtained by adding all the values {$\Gamma_{ijkl}$} located on an anti-diagonal of the JPD defined by {$i-k=i^-$ and $j-l=j^-$ (i.e. $\vec{r_1}-\vec{r_2}=\vec{r^-}$)}. In theory, calculating the minus-coordinate projection of the JPD can therefore achieve pixel super-resolution and potentially retrieved lost spatial information of undersampled objects. However, it is important to note that the JPD anti-diagonal $\Gamma_{ij-i-j}$ and minus-coordinate projection $P^-$ images are not directly proportional to {$|t(x_i,y_i)|$} as in the near-field configuration, but to $|t(x_i,y_i)| |t(-x_i,-y_i)|$, which does not always enable to retrieve $|t|$ without ambiguity. In works using a far-field illumination configuration, this problem is solved by illuminating $t$ with only half of the photon pairs beam (i.e. $t(x,y \leq 0 )=1$). The object then appears twice in the retrieved image (object and its symmetric), but no information is lost. \\
\\
\noindent \textbf{JPD filtering.} Figure~\ref{Figure5}.a shows the sum-coordinate projection $P^+$ calculated using equation~\ref{equ2} from the unfiltered JPD measured in {Figure~\ref{Figure2}}. This image is very noisy and does not reveal the object. To solve this issue, a filtering method is applied and consists in setting to $0$ all values of the JPD that have a negligible weight (i.e. values close to zero) so that their associated noise does not contribute when performing the {sum}. When using a near-field illumination configuration (Figures~\ref{Figure1} and~\ref{Figure2}, and Figure~10 and~15 of the supplementary document), filtering is applied by setting all JPD values to zeros except from those of the main JPD diagonal {$\Gamma_{ijij}$} and of the $8$ other diagonals directly above and bellow i.e. $\Gamma_{ij\,(i\pm 1)\,j}$, $\Gamma_{ij\,i\,(j \pm 1)}$ and $\Gamma_{ij\,(i\pm 1)\,(j \pm 1)}$. When using a far-field illumination configuration (Figures~\ref{Figure3}.a and {Figure~16 of the supplementary document}), filtering is applied by setting all JPD values to zeros except from those of the main JPD anti-diagonal {$\Gamma_{ij\,-i\,-j}$} and those of the $8$ other anti-diagonals directly above and bellow i.e. {$\Gamma_{ij\,(-i \pm 1)\,-j}$, $\Gamma_{ij\,-i\,(-j \pm 1)}$ and $\Gamma_{ij\,(-i \pm 1)\,(-j \pm 1)}$}. diagonal {$\Gamma_{ijij}$} and of the $8$ other diagonals directly above and bellow i.e. {$\Gamma_{ij\,(i\pm 1)\,j}$, $\Gamma_{ij\,i\,(j \pm 1)}$ and $\Gamma_{ij\,(i\pm 1)\,(j \pm 1)}$}. 
 \\
\\ 
\noindent \textbf{Normalisation.} Figure~\ref{Figure5}.b shows the sum-coordinate projection $P^+$ directly calculated from the filtered JPD measured in the experiment in {Figure~\ref{Figure2} before normalisation}. We observe that this image has an artefact taking the form of inhomogeneous horizontal and vertical stripes. This artefact is very similar to this commonly observed in shift-and-add super-resolution techniques~\cite{farsiu_fast_2004} and is often refereed as a 'motion error artefact'. {In our case, it originates from the difference in the effective detection areas of photons pairs when they are detected by the same pixel (diagonal elements) or by neighbouring pixel (off-diagonal elements), as illustrated in Figure~\ref{Figure1bis}}. Using an analogy with the shift-and-add technique, our problem is equivalent to a situation in which the different shifted low-resolution images were measured using {cameras with different pixel width} during the first step of the process. Then, when these low-resolution images are recombined into a high-resolution one (second step of shift-and-add), the artifact appears in the resulting image because some pixels {are less intense than others}. In practice, the simplest way to remove this artifact is to normalize each low-resolution image by its total average intensity so that neighboring pixels in the high-resolution image are at the same level. We use such a normalization approach in our work by dividing all values of the non-zero diagonals in the filtered JPD by their spatial average value i.e. {$\Gamma_{ij\,(i+i^-)\,(j+j^-)} \rightarrow \Gamma_{ij\,(i+i^-)\,(j+j^-)} / \sum_{i,j} \Gamma_{ij\,(i+i^-)\,(j+j^-)}$, where $(i^-,j^-)$} identifies a specific JPD diagonal. After normalization, Figure~\ref{Figure2}.f is obtained. In the case of far-field illumination, the same normalisation is applied to the values of the non-zero anti-diagonals in the filtered JPD i.e. {$\Gamma_{ij\,(i+i^+)\,(j+j^+)} \rightarrow \Gamma_{ij\,(i+i^+)\,(j+j^+)} / \sum_{i,j} \Gamma_{ij\,(i+i^+)\,(j+j^+)}$, where $(i^+,j^+)$}, where $(i^+,j^+)$ identifies a specific JPD anti-diagonal.\\
In some cases the artefact is reduced but still visible in the resulting image even after normalisation. The persistence of this artefact is due the fact that the difference {in the effective integration areas between diagonal and off-diagonal elements is too large to be accurately corrected by simple sum normalization.} {For example, in the experiment shown in Figure~10.a of the supplementary document, the poor quality of the SPAD camera sensor, in particular its very low fill-factor ($10.5\%$), is probably at the origin of the remaining artefact}. To further reduce this artefact, one could use more complex normalisation algorithms, such as $L_1$ or $L_2$ norms minimisation approaches~\cite{farsiu_practical_2006} and kernel regression~\cite{takeda_kernel_2007}, that are commonly used in shift-and-add approaches.   \\
\\
{\noindent \textbf{Slanted-edge method.} MTF measurements using the slanted-edge approach were performed with a razor blade titled by approximatively $100$ mrad and positioned in the object plane of the experimental configuration shown in Figure~\ref{Figure1}, followed by a standard method described in~\cite{burns2000slanted}. Broadening of the curves is estimated by comparing spatial frequency values where MTF is $50\%$ of its low frequency value (i.e. criteria MTF50). More details are provided in section 4 of supplementary document.\\}
\\
{\noindent \textbf{Estimation of the photon correlation width $\sigma$.} \textit{Near-field illumination configuration:} The value of the correlation width in the image plane $\sigma$ is obtained by calculating the position-correlation width at the output of the crystal using the formula $\sqrt{\alpha L \lambda_p / (2 \pi)}$ ($L$ is the crystal thickness, $\lambda_p$ is the pump beam wavelength and $\alpha$ is a parameter described in~\cite{chan_transverse_2007}) and multiplying it by the magnification of the imaging system.\\
\textit{Far-field configuration:} The value of $\sigma$ is obtained by calculating the angular-correlation width of photons at the output of the crystal using the formula $\lambda_p / (2 \omega)$~\cite{chan_transverse_2007}, ($\omega$ is the pump beam waist) and multiplying it by the effective focal length of the imaging system. \\
In our work, values of $\sigma$ are estimated using the theory and not with the experimental techniques described in~\cite{edgar_imaging_2012,moreau_realization_2012} because these approaches fail at providing an accurate result precisely when the correlation width is smaller than the pixel width. In addition, note also that these width values are not strict bounds but correspond to standard deviation widths when modelling spatial correlations with a Gaussian model~\cite{fedorov_gaussian_2009}. More details are provided in section 2.4 of the supplementary document.\\}
\\
\noindent \textbf{Analytical derivation of $\Gamma_{kk}$ and $\Gamma_{kk+1}$.} We consider an unidimensional object $t(x)$ illuminated by photon pairs using a near-field illumination configuration. Photon pairs are characterised by a two-photon wavefunction $\Psi_t(x_1,x_2)$ in the object plane. JPD values $\Gamma_{kl}$ are measured using an array of pixels with pitch $\Delta$ and gap $\delta$. Assuming that the imaging system is not limited by diffraction but only by the sensor pixel resolution, its point spread function can be approximate by a Dirac delta function and $\Gamma_{kl}$ can be formally written as:
\begin{equation}
\label{anadev}
\Gamma_{kl} = \int_{x_k-\frac{\Delta-\delta}{2}}^{x_k+\frac{\Delta-\delta}{2}}  \int_{x_l-\frac{\Delta-\delta}{2}}^{x_l+\frac{\Delta-\delta}{2}} |t(x_1)t(x_2) \Psi_t(x_1,x_2)|^2 dx_1 dx_2,
\end{equation}
where we assumed unity magnification between object and camera planes. A graphical representation of this integral is shown in Figure~\ref{Figure5}.c. For clarity, we only represented an array of three pixels. The bivariate function $|t(x_1)t(x_2)|^2$ is represented in green and overlaps with an grid of squares of size $\Delta$ and spacing $\delta$. Each square represents an integration area associated to a specific JPD value. For example, the central square corresponds to the integration area of $\Gamma_{kk}$ i.e. $[x_k-\frac{\Delta-\delta}{2},x_k+\frac{\Delta-\delta}{2}] \times [x_k-\frac{\Delta-\delta}{2},x_k+\frac{\Delta-\delta}{2}]$. In addition, the bivariate function $|\Psi_t(x_1,x_2)|^2$ is represented by two dashed black lines. These two lines delimit the most intense part of the function, which corresponds to a diagonal band of width $\sigma$ using a double Gaussian model~\cite{fedorov_gaussian_2009}. \\
We seek to calculate the JPD values $\Gamma_{kk}$ and $\Gamma_{kk+1}$. Graphically, these values are located at the intersection between the grid, the green area and the surface inside the dashed lines. They are represented in blue and red, respectively. For small widths $\sigma < \Delta$, it is clear in Figure~\ref{Figure5}.c that the blue and red integration areas are tightening around the positions $(x_k,x_k)$ and $(x_{k+1/2},x_{k+1/2})$ positions, which results in $\Gamma_{kk} \sim |t(x_k)|^4$ and $\Gamma_{kk+1} \sim |t(x_{k+1/2})|^4 $. More formally, one can also apply a change of variable and perform a first-order Taylor expansion in Eq.~\ref{anadev} to reach the same results:
\begin{eqnarray}
\label{anadev2}
\Gamma_{kk} &\approx& |t(x_k)|^4 S_0 \\
\Gamma_{kk+1} &\approx& |t(x_{k+1/2})|^4 S_1 
\end{eqnarray}
where $S_0 \approx \sigma / 2 [1+2 \sqrt{2}(\Delta-\sigma/\sqrt{2})]$ and $S_1 \approx \sigma^2/2$. Full calculations are provided in section 2.5 of the supplementary document. Note also that the difference between integration areas $S_1 \neq S_2$ makes the normalization step after calculating the JPD projection necessary (see Methods section \textit{Normalization}). 

\section*{Aknowledgements} D.F. acknowledges financial support from the Royal Academy of Engineering Chair in Emerging Technology,  UK Engineering and Physical Sciences Research Council (grants EP/T00097X/1 and EP/R030081/1) and from the European Union's Horizon 2020 research and innovation programme under grant agreement No 801060. H.D. acknowledges funding from the European Union's Horizon 2020 research and innovation programme under the Marie Skłodowska-Curie grant agreement No. 840958. JZ has received funding from the European Union's Horizon 2020 research and innovation programme under the Marie Skłodowska-Curie Grant Agreement No. 754354.

\section*{Contributions} H.D conceived the original idea, designed and performed experiments, analysed the data and prepared the manuscript. P.C. performed the slanted-edge experiment. P.C. and H.D. performed the simulations. M.R. and J.F. contributed to developing the quantum illumination protocol with the EMCCD camera. J.Z and E.C. contributed to developing the quantum illumination protocol with the SPAD camera. A.L. and B.N. contributed to developing the entanglement-enabled quantum holography protocol. {A.R.H. participated to the data analysis.} All authors contributed to the manuscript. D.F. supervised the project. \\

\section*{Competing interests.} The authors declare that there are no conflicts of interest related to this article. For the sake of transparency, the authors would like to disclose that EC holds the position of Chief Scientific Officer of Fastree3D, a company which is active in LIDAR and consumer electronics, and that he is co-founder of Pi Imaging Technology. Both companies have not been involved with the paper drafting, and at the time of writing have no commercial interests related to this article.

\section*{Data availability.} The data generated in this study have been deposited in a database under accession code DOI: http://dx.doi.org/10.5525/gla.researchdata.1269.\\
\\
%
\bibliographystyle{naturemag}

\end{document}